%% file: rbsbks.tex
\documentclass{aa}
\usepackage{graphics}

\usepackage{natbib}
\bibpunct{(}{)}{;}{a}{}{,}
\input{thesiscom.tex}
\begin{document}

\title{Reversal-free Ca\,{\sc ii}\,H profiles: a challenge for solar chromosphere modeling in quiet inter-network}
\author{R. Rezaei\inst{1}, J.H.M.J.~Bruls\inst{1}, W.~Schmidt\inst{1}, 
C.~Beck\inst{2}, W.~{Kalkofen}\inst{3} \& R.~Schlichenmaier\inst{1}}
\institute{Kiepenheuer-Institut f\"ur Sonnenphysik, 
Sch\"oneckstr. 6, 79104 Freiburg, Germany  \and
Instituto de Astrof\'isica de Canarias~(IAC), Via L\'actea, E 38\,205, La Laguna, Spain \and
Harvard-Smithsonian Center for Astrophysics, 60 Garden Street, Cambridge, MA 02138
\\
E-mails: [rrezaei,\,bruls,\,wolfgang,\,schliche]@kis.uni-freiburg.de,\,cbeck@iac.es,\,wolf@cfa.harvard.edu}
\date{Received: 12 November 2007/Accepted: 11 March 2008}
\titlerunning{Reversal-free Ca\,{\sc ii}\,H profiles: a challenge for solar chromosphere modeling}
\authorrunning{Rezaei et al.}

\abstract
{}
{We study chromospheric emission to understand the temperature stratification in the solar 
chromosphere.}
{We observed the intensity profile of the \ion{Ca}{ii}\,H line  in a  quiet Sun region 
close to the disk center at the German Vacuum Tower Telescope. 
We analyze over $10^5$ line profiles from inter-network regions. 
For comparison with the observed profiles, we synthesize spectra for a variety of 
model atmospheres with a non local thermodynamic equilibrium~(NLTE) radiative transfer code.
}
{A fraction of about 25\,\% of the observed \ion{Ca}{ii}\,H line 
profiles do not show a measurable emission peak 
in $H_{\mathrm{2v}}$ and $H_{\mathrm{2r}}$ wavelength bands (reversal-free). 
All of the chosen model atmospheres with a temperature rise 
fail to reproduce such profiles. On the other hand, the synthetic calcium profile of 
a model atmosphere that has a monotonic decline of the temperature with height 
shows a reversal-free profile that has much lower intensities 
than any observed line profile.}
{
The observed reversal-free profiles indicate the existence of cool patches in the 
interior of chromospheric network cells, at least for short time intervals. 
Our finding is not only in conflict with a full-time hot chromosphere, but also 
with a very cool chromosphere as found in some dynamic simulations.}

\maketitle

\section{Introduction}

Solar model atmospheres are built to reproduce the observed intensity spectrum  
of the Sun in continuum windows and selected lines~\citep[e.g., the series of models initiated by][]{val81}. 
These models rely on temporally and spatially averaged spectra. Hence, they provide information about the 
average properties of the physical parameters in the solar atmosphere~\citep{linskyavrett}. 
At high spatial, spectral, and temporal resolution, differences from the average profiles arise, 
leading to different interpretations of the atmospheric stratification. 
In this context, there is no general agreement 
on the temperature stratification in the solar chromosphere. 
This is closely related to the fundamental problem of the heating mechanisms in the 
chromosphere. 
\cite{carl_stein_94, carl_stein_97} proposed that apart from short--lived 
heating episodes in the chromosphere its 
time--averaged temperature stratification is monotonically decreasing with geometrical height. 
As noted by \cite{carlsson_etal_97}, there are disagreements between this model and SUMER data. 
Moreover, the cool model was criticized by 
\cite{kalkofen_etal_99} who argued that the chromosphere is always hot and can never reach such 
a low activity state as proposed by Carlsson \& Stein. On the other hand, \cite{ayres_02} 
challenged the arguments of Kalkofen et al. based on observations of CO lines at the solar limb. 
The results of the CO observations were supported by three--dimensional MHD simulations 
that suggested very cool structures in the upper atmosphere 
and implied a thermally bifurcated medium in non--magnetic 
regions \citep{wedemeyer_etal_04, wedemeyer_etal_05}.

\begin{figure*}
\resizebox{\hsize}{!}{\includegraphics{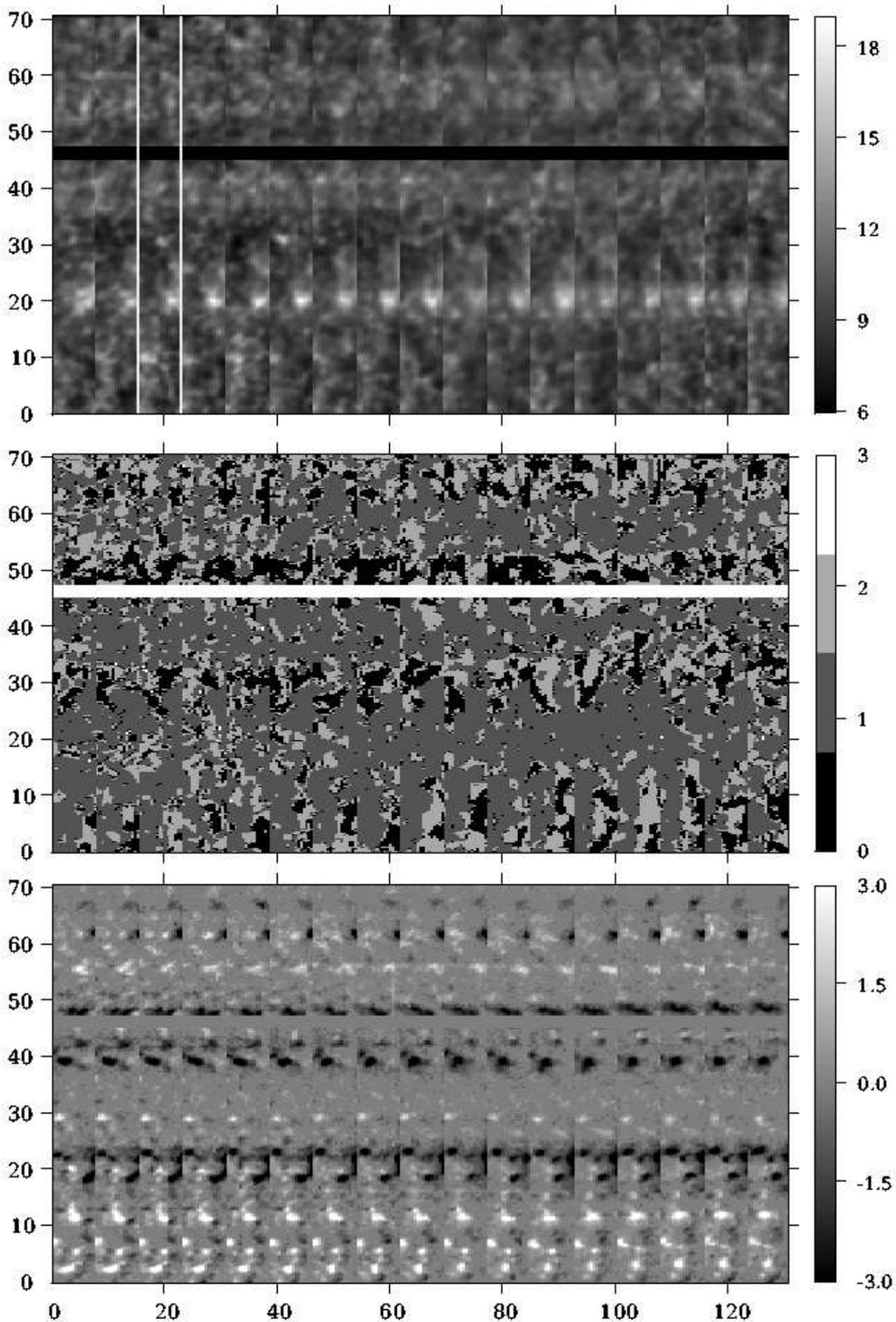}}
\caption[]{Maps of the observed area. \emph{Top:} H-index in pm; \emph{middle:} profile type. 
It is coded as follows: 0: reversal-free; 1: normal; 2: asymmetric; 3: other types. \emph{Bottom:} 
V$_{\mathrm{tot}}$, which is a measure of the net vertical magnetic flux 
density~\citep{lites_etal_99} in arbitrary units. 
The $x$-axis shows both scanning direction and time (width of a single map is eight arcsec). 
The $y$-axis is the slit direction. Both axes are in arcsec. 
Only 17 out of 40 scans are shown here. The time duration 
per scan was about 100\,s.}
\label{fig:map}
\end{figure*}

Recently, \cite{avrett_06a} suggested that it is possible to reproduce a wide range of 
continuum and line intensities with a temperature stratification  
within 400\,K of the semi-empirical models~\citep[][and references therein]{fontenla_etal_06}. 
He concluded that ``...these results appear to conflict with dynamical models that 
predict time variation of 1000 K or more in the chromosphere...''.

In this paper, we comment on both full-time hot and dynamic models, 
based on the lowest intensity profiles we observed. 
We present observations and data reduction in Sect.~2. 
Then, we try to reproduce the observed reversal-free calcium profiles 
(i.e., profiles without emission peaks) by synthesizing line profiles from a variety 
of model atmospheres using a non local thermodynamic equilibrium (NLTE) 
radiative transfer code (Sects.~3 and 4). 
The discussion and conclusion appear in Sects.~5 and 6, respectively.

\section{Observations}
A quiet Sun area close to disk center ($\cos\theta$\,=\,0.99) 
was observed at the German Vacuum Tower Telescope (VTT) in Tenerife, July 07, 2006. 
The good and stable seeing condition during the observation and 
the Kiepenheuer Adaptive Optics System~\citep{luhe_etal_03} provided high spatial resolution in the 
calcium profiles, leading to a collection of different structures observed in the quiet Sun. 
We used an integration time of 4.8\,s and achieved a spatial resolution of 1 arcsec. 
Details of the observation appeared in \cite{reza_etal_4}. 

The \ion{Ca}{ii}\,H intensity profile and Stokes vector profiles of the visible neutral 
iron lines at 630.15\,nm, and 630.25\,nm were observed with the blue and red channels of the 
POlarimetric LIttrow Spectrograph~\citep[POLIS;][]{schmidt03, beck05a}. 
POLIS was designed to provide co--temporal and co--spatial measurements of 
the magnetic field in the photosphere and the \ion{Ca}{ii}\,H intensity profile. 
We normalized the average calcium profile in the line wing at 396.490\,nm to the 
Fourier Transform Spectrograph profile~\citep{stenflo_84} and applied the 
calibration factor to all profiles. Details of the calibration procedure for the 
calcium profiles are explained in \cite{reza_etal_3}. 
The overall properties of the observed time-series are shown in Fig.~\ref{fig:map} concatenated to a 
single map. The H-index, which is the integral of the normalized calcium profile within 
0.1\,nm around the core, is shown in the top panel
~\citep[see][for definitions of the H-index, $V$ and $R$ band intensities]{cram_dame_83,lites_etal_93}. 
The middle panel shows the profile type. It is coded as follows: 
[0] reversal-free, [1] normal (with two emission peaks), 
[2] asymmetric (with only one emission peak), and [3] other types. 
The bottom panel of Fig.~\ref{fig:map} shows V$_{\mathrm{tot}}$, which is a measure of 
the signed magnetic flux~\citep{lites_etal_99}. It is immediately seen that reversal-free profiles are absent 
in the vicinity of network elements ($y\,\sim\,20$\,arcsec in Fig.~\ref{fig:map}).

\begin{figure*}
\resizebox{\hsize}{!}{\includegraphics{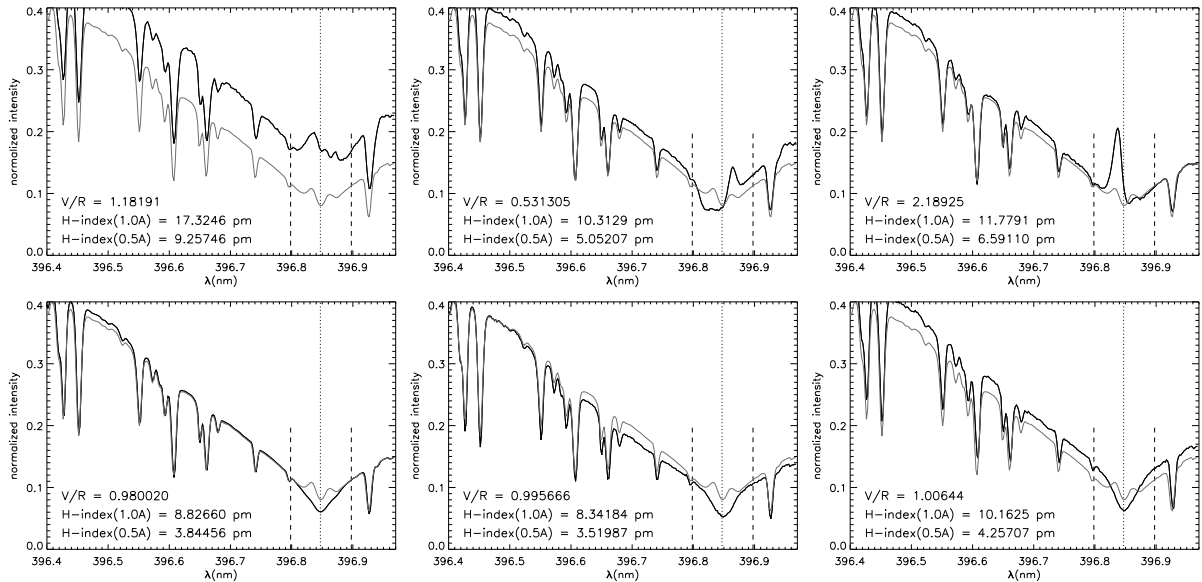}}
\caption{Examples of single calcium profiles with drastically different properties observed in the quiet Sun. 
The gray profile is the average of a few thousands profiles (cell interior and boundary). 
\emph{Top left:} a profile with three emission peaks. 
\emph{Top middle:} a very asymmetric profile with only a red emission peak. 
\emph{Top right:} a very asymmetric profile with two emission peaks. 
\emph{Bottom:} three reversal-free profiles. 
Note the level of the wing intensity with respect to the wing of the average profile. The dashed lines show 
a distance of $\pm$\,0.05\,nm from the average calcium core, indicated by a vertical dotted line. 
The bottom left profile is an average of 59  reversal-free profiles while the other panels 
show single profiles. Except for the top middle and right panels, the V/R ratio was calculated using bands.}
\label{fig:prf}
\end{figure*}

\begin{figure}
\resizebox{\hsize}{!}{\includegraphics{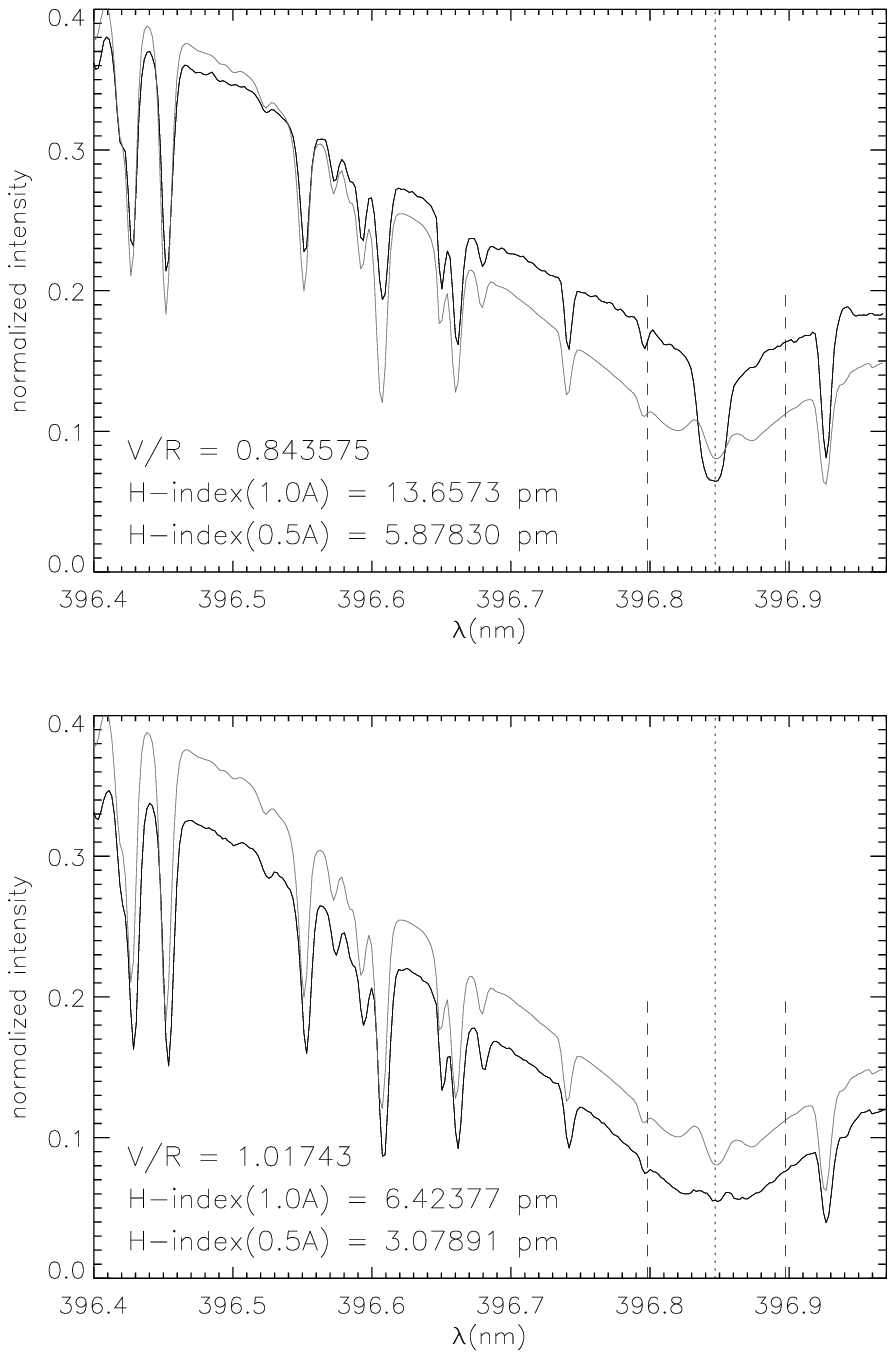}}
\caption{These two examples show that the integrated intensity close to the core, 
either H-index or a filter, is not a reliable criterion to separate reversal-free profiles 
from profiles with emission peaks. Positions (in Fig.~\ref{fig:map}) 
of the upper and lower profile are (18.7, 30.5) and (32.6, 30.2), respectively.}
\label{fig:comp}
\end{figure}

The top row of Fig.~\ref{fig:prf} shows three profiles belonging to 
different activity levels in the quiet Sun. 
The left profile has two red emission peaks, the middle profile 
shows virtually no violet emission peak and the right profile shows a 
strong violet and a weak red emission peak ($V/R > 2.1$). 
The gray profile is the average of a few thousand profiles 
(cell interior and boundary) with an H-index of 10.1\,pm. 
An alternative way to measure the asymmetry of the two emission peaks 
is dividing the peak values rather than band integrated intensities~\citep{reza_etal_3}.
In very asymmetric cases, i.e., where there is only a large violet peak, we can easily achieve 
a large V/R ratio with band definitions whereas the peak ratio is undefined. 
Double reversal calcium profiles in old observations with lower spatial resolution 
do not show extreme values for the ratio of the peaks.  
It requires both high spatial and temporal resolution and large signal to noise ratio. 
The maximum and minimum V/R (peak) ratio in our data set are  2.2 and 0.5 (top right and middle panels, Fig.~\ref{fig:prf}). The importance of using peak values instead of bands is that the formation 
height of different wavelengths in the band can differ by more than one Mm (see Fig.~\ref{fig:cf} 
and the left panel of Fig.~\ref{fig:synth2}). 
The bottom panels of Fig.~\ref{fig:prf} show  calcium profiles observed in absolutely 
quiet Sun regions. There is no emission peak or bulge in the core of these profiles. 
In the bottom row of Fig.~\ref{fig:prf}, the left profile has the same wing intensity as 
the average profile (gray), while the middle and right profiles show lower 
and higher wing intensities than the average profile, respectively. 
Note that the reversal-free profiles usually have a V/R ratio very close to one, 
i.e., they are symmetric (check values in the bottom panels of Fig.~\ref{fig:prf}).
These reversal-free profiles were called absorption profiles by \cite{grossman_etal_74}.
They rightly concluded that the source function of the calcium line  
then shows negligible enhancement in the atmospheric layers where the line forms.
An emission bulge exists in the minimum profile of \cite{cram_dame_83} (their Fig.~\ref{fig:comp}) 
while this is not the case in the bottom panel, Fig.~\ref{fig:prf}. 
This difference means that our reversal-free profiles present a new 
minimum state for the observed \ion{Ca}{ii}\,H profiles. 
Our observations have better spatial and temporal resolutions than \cite{cram_dame_83}, so 
we expect to find more extreme cases than these authors. 

\paragraph{Statistics:} There are many references on statistical properties 
of the \ion{Ca}{ii}\,H line profiles~\citep[e.g.,][]{cram_dame_83},  
but our statistics on reversal-free profiles are the first. 
We exclude the range $y=15-25$\,arcsec in our maps because that clearly contains magnetic network. 
Then we perform a statistical analysis to check how often the reversal-free profiles  occur. 
The inter-network sample has some 129\,000 profiles. 
We check the existence of the emission peaks by analyzing the profile shape. Among the profiles
that have no emission peaks, some show a strong H-index as in 
the top panel of Fig.~\ref{fig:comp}; but others, like the bottom panels of Fig.~\ref{fig:prf}, 
show a rather normal wing intensity and a dark core. 
We find that 25\,\% of all profiles do not show any emission peak:  
16\,\% have an H-index larger than 9\,pm, 8\,\% between 8 and 9\,pm, 
and 1\,\%  below 8\,pm. Examples of calcium profiles with low H-index and no emission peak 
are shown in the lower panels of Fig.~\ref{fig:prf}. 
A majority of these 25\,\% show a tiny bulge at the position of the emission peaks. 

It is worth noting that the H-index alone is not a reliable 
parameter for identifying the reversal-free profiles, as is demonstrated in Fig.~\ref{fig:comp}. 
The upper profile shows no emission peaks, but the wing intensities suggest 
a hotter upper photosphere and a lower photospheric temperature gradient; the 
lower profile has two clear emission peaks, but the wing intensities demand 
a cooler upper photosphere. Consequently, the H-index of the bottom profile 
is less than half of the upper profile. 
This is in accordance with the fact that all three panels of Fig.~\ref{fig:prf} (bottom row) show 
reversal-free profiles  with an H-index larger than 8\,pm.

Besides the 25\,\% fraction of reversal-free profiles in our sample, 
46\,\% of all profiles show a double reversal, with 
strong tendency to have a brighter violet peak. The remaining 29\,\% of all inter-network profiles 
show just one emission peak, again with a majority 
of strong violet emission peaks, similar to previous studies~\citep[e.g.,][]{cram_dame_83}.

\section{Synthetic calcium profiles} 
To reproduce the reversal-free profiles we compute synthetic 
Ca\,{\sc ii}\,H profiles by means of the NLTE radiative transfer code RH, 
developed by \cite{uitenbroek_01}.
The atomic model is the standard 5-level plus continuum model 
that is included with the RH code; apart from minor atomic data updates, 
this model is essentially equal to the one used by \cite{uitenbroek_89}, 
which in turn dates back to \cite{shine_linsky_74} and references therein. 
As shown by \cite{uitenbroek_89}, partial frequency redistribution of 
the photons in both Ca\,{\sc ii}\,H\&K needs to be taken into account, whereas 
assuming complete redistribution in the lines of the infrared triplet 
has no noticeable influence on the profiles of H\&K.
Partial redistribution makes the line source functions wavelength 
dependent and significantly reduces the strength of the H$_{2}$ and 
K$_{2}$ emission w.r.t. the common complete redistribution case. 
Note that since we only deal with static atmosphere models, it suffices 
to use angle-averaged partial frequency redistribution.
Furthermore, \cite{sol_stein91} found that the width of the K$_{2}$  
(and implicitly H$_{2}$) peaks correlates with the microturbulent 
velocities in the chromosphere.
Zero microturbulence produces stronger and narrower peaks whereas 
the microturbulence values given with semi-empirical 1D solar models 
typically result in shallower, wider peaks. 
Microturbulence is a parameter that is used in particular in 1D 
radiative transfer modeling to account for the influence of 
spatially unresolved velocities on the line profiles.
For 2D and 3D dynamic models it is generally not needed.

Since full 2D or 3D (dynamic) radiative transfer modeling is beyond current 
computational resources, we use static 1D plane-parallel models to synthesize Ca\,{\sc ii}\,H 
line profiles, even though the structures that we observe are so small 
that we expect lateral radiative transfer to play a role. 
In addition, we use the microturbulence values as specified with 
the semi-empirical models we employ, even though we do not 
know anything about the unresolved small-scale velocities 
in the observed features.
Synthetic profiles from semi-empirical 1D quiet-Sun models with 
non-zero microturbulence tend to have H$_{2}$-peaks 
that are generally stronger than observed; using zero microturbulence, as one 
would do if the structure to be modeled were completely resolved, 
would only widen this discrepancy. 
Since we cannot be sure that the structure is completely resolved in the 
observations, and also because we expect that radiative interaction 
with the environment will lead to broadening of the H$_{2}$-peaks, 
we decided not to change the microturbulent velocities. 
Note that the latter is not an observational effect, but solely due to 
scattering of line photons in the solar atmosphere. 

\begin{figure*}
\resizebox{\hsize}{!}{\includegraphics{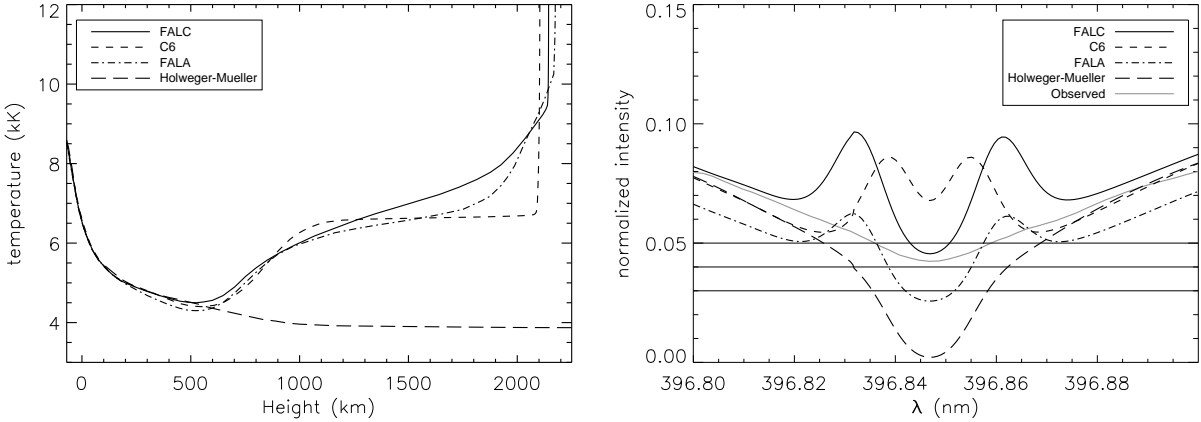}}
\caption[]{\emph{Left:} Stratification of 
the temperature in the FALA, FALC, Holweger-M\"uller, and C6 model atmospheres. 
\emph{Right:} Resulting calcium profiles. 
Except for the Holweger-M\"uller model atmosphere, which does not show a temperature rise, all  
model atmospheres result in two emission peaks. We set the nearby pseudo-continuum 
of the FALC profile to unity and use this normalization factor for all synthetic profiles. 
The observed reversal-free profile in the bottom left panel of Fig.~\ref{fig:prf} is shown in gray. 
All synthetic profiles convolved to the spectral resolution of the observed spectra; 
scattered light removed from the observed profiles.}
\label{fig:synth}
\end{figure*}

As model atmospheres we use four semi-empirical models: FALC (quiet Sun)  
and FALA (inter-network) from \cite{fontenla_etal_99, fontenla_etal_06}; 
the Holweger-M\"uller atmosphere~\citep{hol_mul_74}, which is 
very similar to a theoretical radiative equilibrium model, 
but not constructed to strictly satisfy this condition; 
and the new C6 quiet Sun model atmosphere, provided in advance of 
publication~\citep[][where this model is now called C7]{avrett_loser_07}. 
The left panel of Fig.~\ref{fig:synth} shows the temperature stratification 
of these four model atmospheres. 
All of the models, except the Holweger-M\"uller model, show a chromospheric temperature rise 
at different heights in the atmosphere. The Holweger-M\"uller model has a monotonically-decreasing 
temperature stratification with an artificial extension with constant slope (almost isothermal). 
The extension is required to include the formation height of the Ca\,{\sc ii} resonance lines. 
The synthesized calcium profiles are shown in the right panel of Fig.~\ref{fig:synth}. 
For comparison, the reversal-free profile from the bottom left panel 
of Fig.~\ref{fig:prf} is overplotted in gray.

For all synthesized profiles, except the one derived from the Holweger-M\"uller model, 
there are well-defined emission peaks at the $H_{\mathrm{2v}}$ and $H_{\mathrm{2r}}$ wavelengths. 
All the synthetic profiles are significantly different from the observed reversal-free profiles. 
It might suggest that, e.g., the temperature rise in the chromosphere is either shifted 
to higher layers, or the temperature gradient is not 
as steep as proposed by, for example, the C6 model. For a very inhomogeneous model atmosphere, the 
region with high emission may be very ``patchy''.

The new chromospheric model C6, like FALC, 
includes a chromospheric temperature rise. If one considers a velocity stratification 
in the atmosphere, it will be possible to reproduce a variety of asymmetric calcium 
profiles~\citep[e.g.,][]{heasley_75}. However, it is not trivial to diminish 
both emission peaks at the same time to reproduce a reversal-free profile.
As seen in Fig.~\ref{fig:synth}, none of the applied models with a 
chromospheric temperature rise produces a reversal-free calcium profile. 
While the C6, FALA, and FALC models look too hot, with strong emission peaks, 
the Holweger-M\"uller model fails due to its very deep calcium core. 
The dynamical models of 
\cite{carl_stein_94,carl_stein_97} and \cite{rammacher_cuntz} 
have  episodes (significantly) cooler than the Holweger-M\"uller 
model with correspondingly lower line core intensities \citep{uitenbroek_02,rammacher_05}. 
It should be remembered, however, that 
the chromospheric radiative transfer in those simulations is not ``consistent'': important cooling agents 
are not represented at all and deviations from various equilibrium states are ignored.

\begin{figure}
\resizebox{\hsize}{!}{\includegraphics{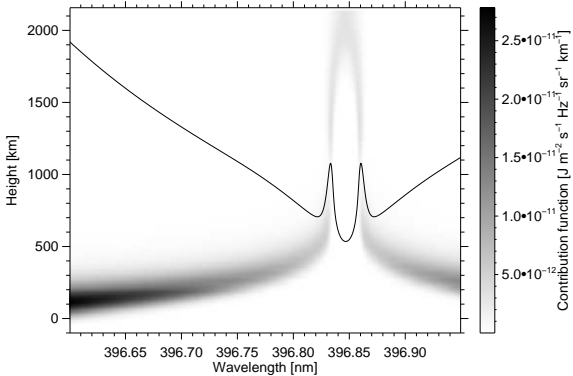}}
\caption[]{Intensity contribution function for the inner part of 
the Ca\,{\sc ii}\,H line computed for the FALC model. 
The black line is the calcium profile of the model.}
\label{fig:cf}
\end{figure}

\section{Modified model atmospheres}
Given the large deviations between all synthetic profiles of the Ca\,{\sc ii}\,H line 
and the observed reversal-free profiles, it is tempting to try to obtain a closer match 
to the observed profiles 
by making changes to one of the model atmospheres, specifically to the FALC model.  
This procedure is by no means unique and given that we employ static models it is 
not clear whether we can reproduce the observed profiles that result from the 
dynamic and highly-structured solar atmosphere. 
We could end up adjusting all atmospheric parameters and still 
not get a satisfactory match.
Instead, we choose to adjust only the FALC temperature in a very schematic way to get 
at least an indication which changes might be needed. 
We stress, however, that these modified models are no longer consistent since only 
the temperature is changed and all other parameters are left intact. They may not 
reproduce observed profiles of other lines. 
From the NLTE radiative transfer computations we know that the monochromatic 
source function follows the local temperature fairly well throughout most of the 
Ca\,{\sc ii}\,H line, even though  it is a strong, scattering line. 
This means that a naive change of the temperature stratification at some height 
translates into a change in emergent intensity for those parts of the line that 
form at that particular height. 
The intensity contribution function for the Ca\,{\sc ii}\,H line 
(Fig.~\ref{fig:cf}) indicates that the formation range for any given wavelength 
is very limited, which means that temperature corrections at a given height will 
lead to predictable line profile changes.
This procedure should work particularly well for the 
line wings up to the H$_1$ minima, where the source function is very close to LTE, 
and for the very core, whose source function drops way below the Planck function 
in a predictable way.  
However, even though the formation range at any wavelength within the H$_2$ peaks is 
narrow, the wide range of formation heights spanned by these peaks as a whole 
necessitates temperature changes over a wide height range to suppress 
the emission peaks completely.  
In addition, we need a rather high temperature in the upper chromosphere 
to obtain a reasonable core intensity; for that reason we use the FALC model as starting point. 
This means that we may only change the temperature structure in the lower and 
middle chromosphere: we created models with the temperature minimum 
region isothermally extended upward to column mass values of $10^{-3.0}$ (ISO\_30), 
$10^{-3.5}$ (ISO\_35), and $10^{-4.0}$ (ISO\_40), followed by a linear 
$T(\log m)$ relation up to a column mass of $m=2\cdot 10^{-5}$\,g/cm$^2$, 
above which the original FALC temperatures are retained (Fig.~\ref{fig:synth2}, left panel). 
The lower temperature in the height range about 600--1900\,km 
weakens the emission peaks, while the 
temperature rise in the higher layers prevents a very deep calcium core. 
The profiles (Fig.~\ref{fig:synth2}, right panel) suggest that a temperature rise 
shifted to larger heights is a step toward producing a reversal-free profile.

\section{Discussion} 

\begin{figure*}
\resizebox{\hsize}{!}{\includegraphics{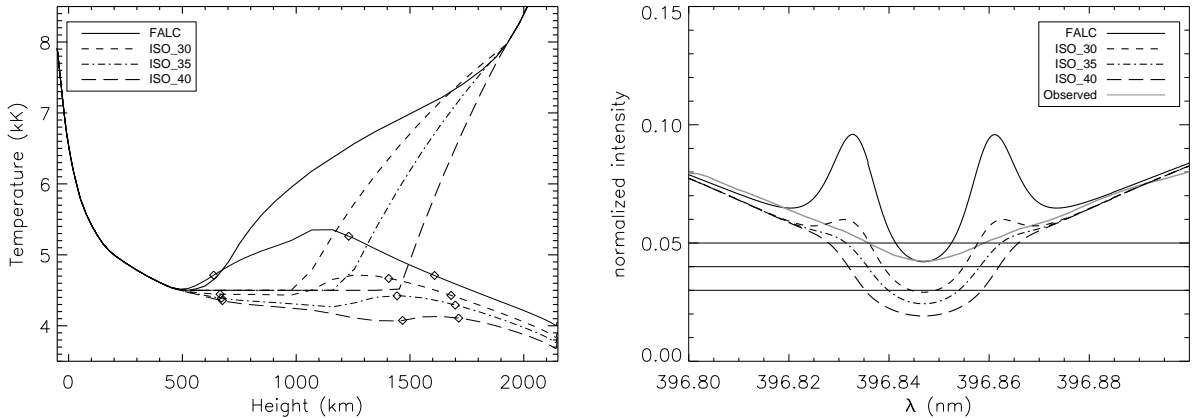}}
\caption[]{\emph{Left:} Stratification of 
the temperature in the FALC model atmosphere and three modified versions of that. The thin curves represent 
the monochromatic source functions of these models at $\Delta \lambda = - 0.014$\,nm from line center. 
The diamonds (from left to right on each curve) represent the location of 
$\tau_{\nu}\,=\,1$ for $\Delta \lambda = - 0.016$, $\Delta \lambda = - 0.014$, and 
$\Delta \lambda = - 0.012$\,nm. The source function is similar for all three wavelengths, so that 
we restrict to just one source function curve per model. 
\emph{Right:} Resulting calcium profiles of the mentioned models convolved to the spectral 
resolution of POLIS. The ISO$\_$30 profile is much closer to the observed 
reversal-free profile than the FALC model.}
\label{fig:synth2}
\end{figure*}

We find that a quarter of all observed profiles are reversal-free profiles. 
We observed reversal-free profiles  of the \ion{Ca}{ii}\,H line 
with an exposure time of about 4.8\,s. However, they may get lost at 
longer exposure times. The maximum spatial extension of these 
reversal-free profiles  in our map is about 5\,arcsec; 
they mostly appear in smaller structures (Fig.~\ref{fig:map}, middle panel). 
The coherency of the position of the reversal-free profiles shows that 
they are not an artifact due to effects caused in the Earth atmosphere.
The observed reversal-free \ion{Ca}{ii}\,H profiles (Fig.~\ref{fig:prf}, lower panels) 
suggest that there may be some cool patches in the chromosphere for short periods of time. 
In contrast, \cite{carlsson_etal_97} tried to find reversal-free profiles in chromospheric 
spectral lines observed with SUMER (46.5--161\,nm). 
On the basis of a few lines they concluded that ``all chromospheric lines 
show emission above the continuum everywhere, all the time''. These lines 
cover a large height range above the classical temperature minimum and should also reveal the 
cool patches seen in \ion{Ca}{ii}\,H line profiles if observed at the proper spatial and 
temporal resolution.

We propose that at a spatial resolution of one arcsecond, 
the average temperature stratification in the solar atmosphere is probably 
hotter than the Holweger-M\"uller model atmosphere, but still does not 
show a permanent temperature rise at all spatial positions.
Moreover, the residual core intensity of these profiles is about 0.05\,I$_{\mathrm{c}}$. 
The Holweger-M\"uller model, as well as comparable cool models 
(like model Q of \citet[][]{sol_stein91} or model 
COOLC of \citet{ayres_etal_86}) and the cool phases of the dynamic models 
 result in a very dark calcium core~\citep{uitenbroek_02}, 
far below the observed reversal-free profiles. 
In these models, we have a longer cool phase and a shorter hot phase 
in the chromosphere~\citep{rutten_98,judge_peter_98}. Statistics of 
our inter-network sample indicates that some 75\,\%  of all profiles 
show one or more clear emission peaks. 
This appears to be in conflict with these models. But we recall that even if at any 
given point in the atmosphere the temperature is low most of the time, the \ion{Ca}{ii}\,H 
line may still show emission peaks most of the time because it is an integral property with 
contributions from a large height range. We speculate that this is even more important in 3D 
models, where lateral interaction with the ubiquitous shocks plays a role as well. 
Unfortunately, so far no one has attempted to compute \ion{Ca}{ii}\,H  profiles simultaneously with the 
3D (M)HD simulations. Even worse, computing \ion{Ca}{ii}\,H a posteriori for a single 
3D snapshot already presents an unsolved task.

On the other hand, our statistical studies of the inter-network calcium profiles indicate that 
about 25\,\% of the profiles do not show any emission peak (but might show a bulge). 
These reversal-free profiles  (bottom panels, Fig.~\ref{fig:prf}) are not reproduced 
by any of the hydrostatic model atmospheres with a chromospheric 
temperature rise, like in the new C6 model. 
This challenges~\cite{avrett_loser_07}, who claim that they can reproduce all 
continuum and line intensities with models that have a temperature variation of 
at most 400\,K. 
In addition, the observed reversal-free profiles  cannot be derived from  
the dynamical models during the cool phase~\cite[][Fig.~6]{uitenbroek_02}. 
Similarly, \cite{avrett_85} tried to modify the temperature stratification of VALA, VALC 
and VALF models to reproduce minimum and maximum profiles of \cite{cram_dame_83}. 
While we kept the temperature minimum constant and extended it isothermally, they introduced a 
hotter temperature minimum and decreased the temperature gradient (their Fig.~17). 
Our model ISO$\_$30 and the corresponding profile (Fig.~\ref{fig:synth2}) better 
fits the observed reversal-free profile, which has lower intensity 
at the emission peak wavelengths than the minimum profile of \cite{cram_dame_83}.

\cite{fontenla_etal_07} presented a compromise solution for the cool patches. 
They proposed that cool patches may occur in a limited height range, such that 
they contribute substantially to the temperature sensitive molecules, 
but affect other spectral lines very little. 
This argument may apply to high resolution filtergrams~\citep{friedrich_etal_06}. 
If we had cool temperatures in a very limited height range 
as proposed by \cite{fontenla_etal_07}, and a hot atmosphere elsewhere, 
we would see emission features somewhere in each \ion{Ca}{ii}\,H spectrum, which is not the case. 
Therefore, although we have strong indications of cool patches in the solar chromosphere, 
they are more extended (in height) than was proposed by \cite{fontenla_etal_07}.

\cite{bala_01} presented multiple reversal calcium profiles and interpreted 
them as a signature of magnetic flux emergence. We observe some calcium profiles 
with more than two emission peaks (Fig.~\ref{fig:prf}, upper left panel) in 
the quiet Sun, without signature of magnetic flux emergence in the polarimetric 
channel of POLIS~\citep[see][for a description of this channel]{beck_etal_07b}. 
Therefore, his explanation is not unique.

\section{Conclusion}
We find that a quarter of the observed calcium profiles in the quiet Sun inter-network 
show a reversal-free pattern. We interpret these profiles as strong indication for 
a temperature stratification cooler than that of the average quiet Sun. 
Although they do not imply a very cool atmosphere, like 
the one based on CO line observations~\citep{ayres_02}  and numerical 
simulations~\citep[e.g.,][]{carlsson_etal_97,rammacher_cuntz}, it should be 
clearly cooler than was suggested by \cite{avrett_loser_07}. 
The reversal-free profiles  we observe, at a spatial resolution of 1\,arcsec and a temporal 
resolution of 5\,s, show a residual intensity at the core of about 0.05\,I$_{\mathrm{c}}$.  
Hence, the profiles suggest that the cool components are cooler than the FALA, FALC, 
and C6 models in the sense that their corresponding profiles do not have any 
emission peak, but hotter than the extended Holweger-M\"uller model or 
the cool phase in the dynamic models.

\begin{acknowledgements}
We thank Han Uitenbroek for providing his
flexible non-LTE radiative transfer code RH. 
We are grateful to Oskar Steiner and Reiner Hammer for a careful reading of the manuscript. 
We also thank the referee for his critical comments that led to 
significant improvements of the manuscript. 
The POLIS instrument has been developed by the Kiepenheuer-Institut in 
cooperation with the High Altitude Observatory (Boulder, USA). 
Part of this work was supported by the Deutsche Forschungsgemeinschaft (SCHM 1168/8-1).
\end{acknowledgements} 

\bibliography{rbsbks}

\end{document}

%% file: thesiscom.tex
\DeclareRobustCommand{\ion}[2]{%
\relax\ifmmode
\ifx\testbx\f@series
{\mathbf{#1\,\mathsc{#2}}}\else
{\mathrm{#1\,\mathsc{#2}}}\fi
\else\textup{#1\,{\mdseries\textsc{#2}}}%
\fi}

\def\apjl{ApJ}%
\def\apjs{ApJS}%
\def\ao{Appl.~Opt.}%
\def\aap{A\&A}%
